# Temperature enhanced photothermal cooling of a micro-cantilever

Hao Fu[1], Li-ping Ding[1,2], Tian-hua Mao[1,2], and Gengyu Cao[1*]

[1] State Key Laboratory of Magnetic Resonance and Atomic and Molecular Physics, Wuhan Institute of Physics and Mathematics, Chinese Academy of Sciences, Wuhan 430071, China

[2] University of Chinese Academy of Sciences, Beijing 100049, China

*Corresponding author: Tel: +86 27 8719 7737; Fax: +86 27 8719 8576.

E-mail address: gycao@wipm.ac.cn (Gengyu Cao)

**Abstract**

We present a temperature enhanced photothermal cooling scheme in a micro-cantilever based FP cavity. Experiments at various temperatures show a temperature dependence of photothermal cooling efficiency. And approximate one order of improvement on the cooling efficiency is achieved experimentally when the temperature decreases from 298 K to 100 K. Numerical analysis reveals that the dramatic change of the cooling efficiency is attributed to the temperature dependent dynamics of the photothermal backaction. A high efficient cooling can be achieved by controlling the temperature for an optimized the dynamics of photothermal backaction.

# 1. Introduction

Dynamical control of mechanical resonator is of highly interesting for both fundamental physics and applied sciences [1-8]. Motivated by observing quantum phenomenon at macro-scales, optomechanical cooling has demonstrated its strong capacity on suppressing the Brownian motion of mechanical resonator by introducing an additional optical damping to the mechanical systems [9-12]. And the milestone work of radiation pressure cooling of mechanical resonator down to its mechanical ground state has been achieved in resolved-sideband limit [13]. Besides the great achievement radiation pressure cooling, photothermal mechanism also attracts intensive attentions in many applications for its relatively low technical requirements [14, 15]. By constructing a low finesse cantilever-based Fabry-Pérot (FP) cavity, pioneering work has demonstrated photothermal cooling of the fundamental mode of a gold coated cantilever down to 18 K from room temperature [16]. And further research on the higher-orders mechanical modes involved photothermal coupling has provided an insight on overcoming the cooling limit imposed by optomechanical instability of the higher-orders modes [17]. In pursuing a better cooling effect, however, other effects such as optical power absorption and optomechanical bistability may limit the result of this effort at stronger coupling condition [18]. Therefore, in order to improve the cooling limit, it is of critical importance to realize a high efficient photothermal cooling.

In this paper, we studied the photothermal cooling in a low finesse cantilever-based FP cavity at various temperatures. Experimental results show that the cooling efficiencies are dramatically different at temperatures range from 298 K to 78 K. Further analysis reveals that the change of the photothermal cooling efficiency originates from temperature dependent dynamics of the photothermal backaction. And almost one order of improvement on the photothermal cooling efficiency has been achieved experimentally by operation at the optimal temperature of 100 K.

## 2. Experiment

The experimental setup is schematically illustrated in Fig. 1. A compliant 300 μm×10 μm×0.85 μm single crystal silicon micro-cantilever is used to construct a low finesse FP cavity with a plane fiber. The whole setup is immersed into a liquid nitrogen cryostat, where the temperature can be precisely regulated by a temperature controller (Lakeshore 340) from 78 K to 300 K with a stability of ±0.1 K [14]. Gold films of 50 nm thick are deposited on both sides of the cantilever to enhance the photothermal coupling and to prevent stress induced bending of cantilever at cryogenic condition at meanwhile. To avoid air viscous damping, the cavity optomechanical system is placed in an ultra-high vacuum chamber with a base pressure better than $5\times10^{-10}$ Torr. At room temperature condition, the intrinsic resonant frequency and damping factor of the cantilever are $2\pi\times9897.05$ Hz and $2\pi\times4.14$ Hz respectively. The oscillation of cantilever is measured by a 1310 nm laser interferometer and analyzed in frequency domain using an FFT spectrum analyzer (SR760, Stanford Research System). The cavity resonance is tuned by controlling the fiber position via a piezo. The photothermal cooling is performed at blue detuning point *b*, indicated in the inset of Fig. 1.

## 3. Results

Mechanical resonances of the cantilever measured for different laser powers at room temperature are shown in Fig. 2(a). The amplitude of Brownian motion of the cantilever decreases gently as the laser power increasing. And only 7% increase of the effective damping factor is observed when the laser power $P$ is increased from 34 μW to 419 μW. However, as shown in Fig. 2(b), when the setup is immersed into a liquid nitrogen cryostat, the effective damping factor $\Gamma_{eff}$ increases from $2\pi \times 0.57$ Hz to $2\pi \times 4.74$ Hz as the laser power increases from 21 μW to 316 μW. As comparing to that at room temperature, photothermal backaction exhibits much stronger capacity on suppressing the Brownian motion of the cantilever at 78K.

The temperature dependence of photothermal cooling is further investigated at various temperatures range from 78K to 298K. The effective damping factors of the cantilever at five temperatures are plotted as function of laser power in Fig. 3. Although the effective damping factors increases linearly with laser power increasing, the cooling efficiencies varies obviously at different temperatures. While a 100 μW increase of laser power results in the effective damping factor increasing $2\pi \times 1.36$ Hz at 78 K, it cases only $2\pi \times 0.44$ Hz and $2\pi \times 0.15$ Hz increasing of $\Gamma_{eff}$ at 160 K and 298 K respectively.

The photothermal cooling efficiency, which is defined as $\eta_{ph} = d\Gamma_{eff}/dP$, is analyzed quantitatively by linearly fitting the curves. As demonstrated in Fig. 4, the photothermal cooling efficiency is improved gradually when temperature changes from 298 K to 150 K. As the temperature further decreasing to 100 K, a significant enhancement is observed with the photothermal cooling efficiency increasing 9.7 times from 2π×1.5 mHz/μW to 2π×14.5 mHz/μW. Operation below the temperature of 100 K, a slightly decreasing of the photothermal cooling efficiency is observed.

## 4. Discussion

Cooling by this cold damping technique, the Brownian motion of the cantilever is suppressed by introducing an additional optical damping $\Gamma_{opt}$ to modify the mechanical damping as $\Gamma_{eff}=\Gamma_0+\Gamma_{opt}$. The optical damping $\Gamma_{opt}=\chi(\omega)g(P)$ is a product of the dynamical response of the photothermal force $\chi(\omega)$ at frequency ω and the strength of the backaction g(P), which is proportional to the laser power. While the cooling efficiency can be improved by employing high quality optical cavities and mechanical resonators to provide a strong coupling, the efficiency can also be enhanced by optimizing the dynamical response of photothermal force. For a mechanical mode with resonant frequency $\omega_m$, the response function can be written as $\chi(\omega)=\omega_m\tau_{ph}/(1+\omega^2\tau_{ph}^2)$, where $\tau_{ph}$ is response time constant of photothermal force to the mechanical motion [19].

The retard behavior of photothermal force is of critical importance for cooling operation. An instantaneous photothermal backaction with $\tau_{ph} \ll 1/\omega_m$ results in $\chi(\omega_m)\approx0$, which indicates a extremely weak cooling. However, on the other hand, a slow response with $\tau_{ph} \gg 1/\omega_m$ also contributes little to the cooling operation for the backaction of photothermal force is averaged out for a oscillation period. An optimal photothermal cooling can be realized at the condition of $\omega_m\tau_{ph}=1$.

The response of the photothermal force is directly related to the time constant of heat diffusion along the cantilever, which is determined by many temperature-dependent parameters such heat capacity and thermal conductivity of the materials [19, 20]. The temperature dependent dynamics of the photothermal backaction offers a possibility to enhance the photothermal cooling efficiency by controlling the temperature. Numerical calculation shows that only a weak cooling effect can be obtained at 298 K for $\omega_m\tau_{ph}(298K)=2.5$. However, as temperature decreased to 100 K, the calculation result of $\omega_m\tau_{ph}(100K)=1.04$ indicates that the optimal cooling condition is approximated and hence a high efficient cooling achieved experimentally. Fitting of the experimental measured photothermal cooling efficiency reveals that the strength of photothermal backaction is $dg(P)/dP=2\pi\times 22.3$ mHz/μW. The deviation of experimental results from theoretical calculation at low temperatures in Fig. 4 could be attributed to temperature induced photothermal coupling strength changes. For example, cantilever deformation as well as the change of material stress and thermal expansion coefficient cased by temperature change can affect the strength of photothermal coupling [91].

## 5. Conclusions

In conclusion, we have demonstrated that the dynamics response of the photothermal force can be optimized for a high efficient photothermal cooling by controlling the environmental temperature. Experimental results of photothermal cooling of the gold coated cantilever at various temperatures show a temperature dependent behavior of the cooling efficiency. With the decreasing of temperature from 298 K to 78 K, an optimal photothermal cooling is achieved at 100 K with approximate one order of improvement on the photothermal cooling efficiency as comparing to that at 298 K. Numerical results reveal that the enhancement of the cooling efficiency originates from the temperature dependent nature of the dynamics of the photothermal backaction. When the dynamics of the photothermal backaction is modified such that $\omega_m\tau_{ph}=1$ is satisfied, optimal photothermal cooling can be achieved. Therefore, while the optomechanical cooling efficiency can be enhanced by improving the optomechanical coupling strength, we conclude that a higher cooling efficiency can be benefited from optimizing the dynamics of optical force further. Along with optimizing the dynamics of photothermal force, we note here that the optimal cooling condition can also be satisfied by controlling the mechanical resonant frequency alternatively.

**Acknowledge**

This work was supported by the Grand Project of Instrumentation and Equipments for Scientific Research of the Chinese Academy of Sciences under Grant No. YZ0637 and National Natural Science Foundation of China under Grant No. 11204357.

**Figures and captions**

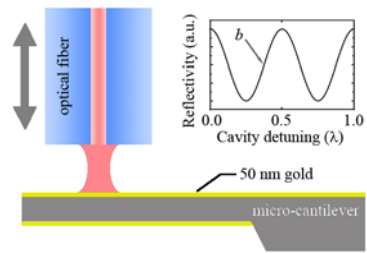

Figure 1. Schematic of the experimental setup. Laser spot is positioned at 150 μm to the fixed end of cantilever. Inset: reflectivity of the micro-cantilever based FP cavity measured by changing the fiber position. Photothermal cooling is performed at the maximum interferential slope point *b*.

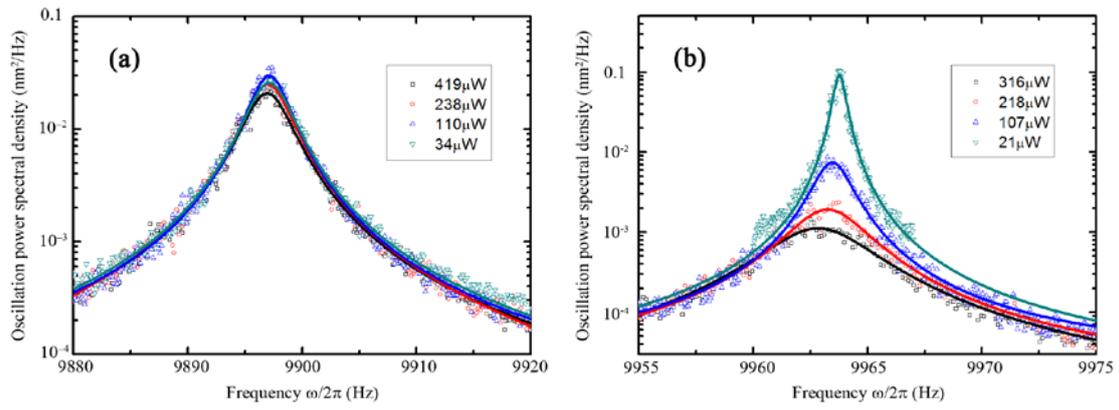

Figure 2. Oscillation power spectral densities of the fundamental mechanical mode cantilever measured at (a) 298 K and (b) 78 K. The vibration resonance curves are Lorentzian fitted (solid lines).

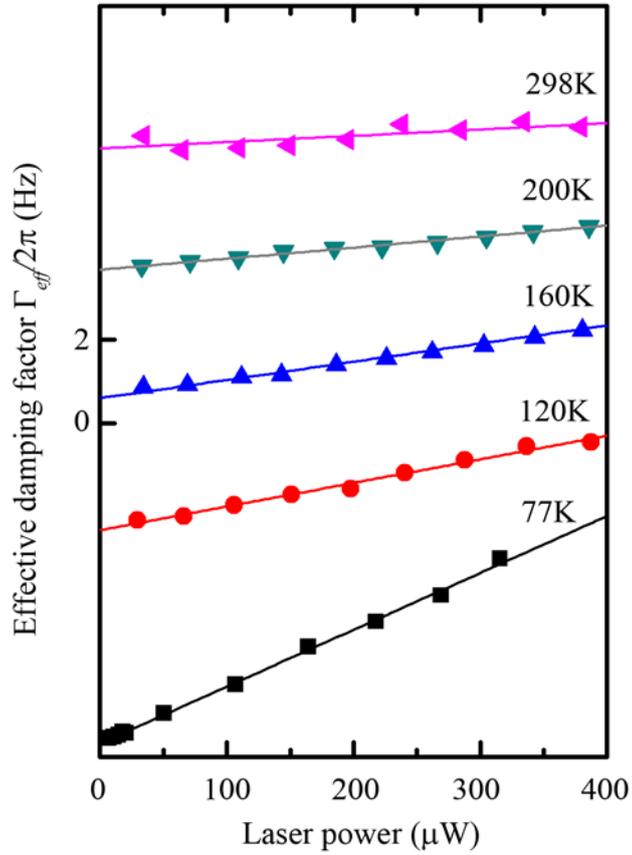

Figure 3. Effective damping factor as a function of laser power. Experimental measurements at five temperatures (dots) are linearly fitted (solid lines) to obtain the photothermal cooling efficiency.

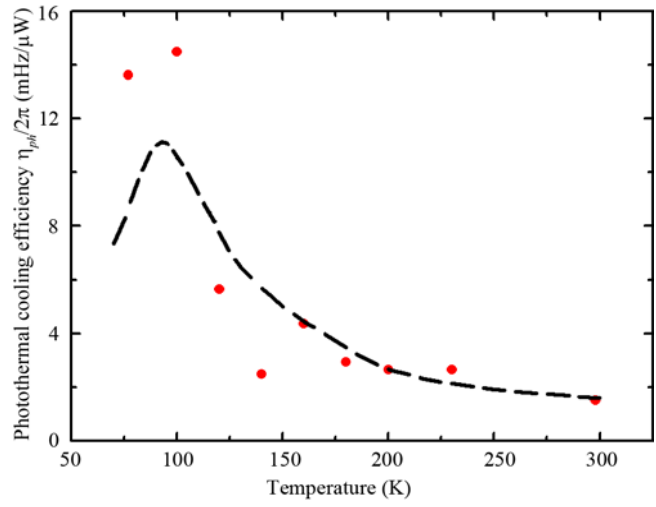

Figure 4. Photothermal cooling efficiency at different temperatures. Theoretical calculation (dash line), which reveals that optimal photothermal cooling appears at 95 K, agrees well with the experimental results (dot).